\documentclass[usenatbib]{mn2e}
\usepackage{graphicx,times}

\usepackage{amsmath,amssymb}

\begin{document}


\title[Tidal Imprints Of A Dark Sub-Halo]{Tidal Imprints Of A Dark Sub-Halo On The Outskirts Of The Milky Way}

\author[Sukanya Chakrabarti \& Leo Blitz]{Sukanya Chakrabarti$^{1,2}$ \& Leo Blitz $^{1}$\thanks{Email: sukanya@astro.berkeley.edu}\\
$^1$601 Campbell Hall, Astronomy Dept., University of CA Berkeley, Berkeley, CA 94720, sukanya@astro.berkeley.edu\\
$^2$UC Presidential Postdoctoral Fellow}


\maketitle

\begin{abstract}
We present a new analysis of the observed perturbations of the HI disk of the Milky Way to infer the existence of a dark sub-halo that tidally interacted with the Milky Way disk.  We examine tidal interactions between perturbing dark sub-halos and the gas disk of the Milky Way using high resolution Smoothed Particle Hydrodynamics (SPH) simulations.  We compare our results
to the observed HI map of the Milky Way to find that the Fourier
amplitudes of the planar disturbances are best-fit by a perturbing
dark sub-halo with a mass that is one-hundredth of the Milky Way with
a pericentric distance of 5 kpc.  This best-fit to the Fourier modes occurs about 
a dynamical time after pericentric approach, when the perturber is 
90 kpc from the galactic center.  Our analysis here represents a new method to indirectly characterize dark sub-halos from the tidal gravitational imprints they leave on the gaseous disks of galaxies.  We also elucidate a fundamental property of parabolic orbits.  We show that under certain conditions, one can break the degeneracy between the mass of the perturber and the pericentric distance in the evaluation of the tidal force -- to directly determine the mass of the dark perturber that produced the observed disturbances.
\end{abstract}

\section{Introduction}

Recent HI maps of the Milky Way display perturbations with amplitudes of order unity well outside the solar circle (Levine, Blitz \& Heiles 2006).  The strength and scale of these perturbations cannot be expected in an isolated galaxy where the gas responds to forcing by the stellar spiral arms.  In this $\it{Letter}$, we explore the possibility that these disturbances arise from passing dark sub-halos as they tidally interact with the Milky Way disk.  We are motivated by the paradigmatic view of structure formation in the cold dark matter model (White \& Rees 1978), wherein structure grows hierarchically, with small objects merging to form larger bodies.  Numerical simulations suggest an abundance of cold dark matter substructure (Moore et al. 1999; Klypin et al. 1999; Springel et al. 2005; Diemand et al. 2008 and references therein), prompting an analysis of the tidal imprints of passing dark sub-halos on the Milky Way disk.  In isolated galaxies, density waves induced by the stellar spiral arms would not significantly affect the gas outside the outer Lindblad resonance, as demonstrated in recent self-consistent high-resolutions simulations (Chakrabarti 2009).  The observations of Levine et al. (2006) therefore raise the question -- what mechanism induced these structures in the Milky Way disk?

The morphological signatures of the tidal interactions of dark sub-halos (or mini-halos) and stellar disks have recently been explored (Kazantzidis et al. 2008; Bekki 2009).  The response of a given component of a galaxy scales inversely with the square of its effective sound speed.  Thus, dark sub-halos would more pronouncedly affect the cold gas in a galaxy than the stellar component, in proportion to the ratio of the square of their effective sound speeds.  This suggests that the probability of detecting dark matter sub-halos via analysis of their tidal imprint on the galactic disk would be maximized by looking for perturbations in the cold gaseous components of galaxies, i.e., in the extended HI disks.  

While the over-abundance of sub-structure in simulations relative to observations of Milky Way satellites has received much attention and has been dubbed the ``missing satellites problem'', it is important to underscore the relative importance of the most massive sub-structures in gravitationally sculpting the gas disks of galaxies.  In the impulsive heating approximation, the tidal effects of a sub-halo population scale as $dE/dt \propto \int n(M_{\rm sat})M_{\rm sat}^{2}dM_{\rm sat}$, where $n(M_{\rm sat})$ and $M_{\rm sat}$ are the number density and satellite mass (White 2000; Kazantzidis et al. 2008).  Cosmological simulations predict that the mass function of sub-halos is given by a power-law $n(M_{\rm sat}) \propto M_{\rm sat}^{-\alpha}$, with $\alpha \sim 1.8-1.9$ (Gao et al. 2004).  Therefore, we can expect that the dynamical effects of CDM sub-structures will be dominated by the most massive sub-structures similar to Large Magellanic Cloud (LMC) sized objects, rather than the structures that are on the rising tail of the dark sub-halo mass function.  

We employ high resolution hydrodynamical simulations in this paper to study the tidal effects of dark sub-halos on the the Milky Way.  Aside from the tidal imprint that dark sub-halos leave on the galactic disk, these objects are unlikely to be detectable via other means, such as lensing, due to their low surface density on the sky.  Thus, it is particularly important to analyze their tidal footprints in the observed HI map to arrive at an independent measure of the masses and pericentric approach distances of the sub-halo population.   Towards this end, we carry out controlled numerical experiments by performing an exhaustive parameter space survey where we vary all the relevant parameters of a sub-halo population, including the masses of
dark satellites, their pericentric approach distance, and orbital
inclination.  We also vary parameters of the primary galaxy that is designed to simulate the Milky Way, such as the equation of state of the gas, gas fraction, and circular velocity.  These details are deferred to \S 2.  We compare the Fourier modes of the planar disturbances of the simulations and the observations; the simulation that minizes the residuals is then taken to be our best-fit simulation.  To address the potential degeneracy between mass and pericentric distance in the evaluation of the tidal force, $F_{\rm tide} \propto M_{\rm sat}/R_{\rm peri}^{3}$, we elucidate the physical conditions under which one can discriminate between tidal interactions of satellites of varying mass at their equivalent tidal distances in \S 3.  Salient points are summarized in \S 4.

\section{Simulation Methodology}

We employ the parallel TreeSPH code GADGET-2 (Springel 2005) to perform simulations of disk galaxies tidally interacting with dark matter mini-halos.  GADGET-2 uses an N-body method to follow the evolution of the collionsionless components, and SPH to follow the gaseous component.  The simulations reported here (unless otherwise noted) have gravitational softening lengths of $100~\rm pc$ for the gas and stars, 
and $200~\rm pc$ for the halo.  The number of gas, stellar, and halo particles in the primary galaxy are $4 \times 10^{5}$, $4 \times 10^{5}$ and $1.2 \times 10^{6}$ respectively for our fiducial case.  

The halo of the primary is initialized with a Hernquist (1990) profile with an effective concentration of 9.4, a spin parameter $\lambda=0.036$, and a circular velocity $V_{200}=160~\rm km/s$.  The primary galaxy is by construction designed to be similar to the Milky Way.  Therefore, we include an exponential disk of stars and gas, with a flat extended HI disk, as found in surveys of spirals (e.g. Wong \& Blitz 2002).  The exponential disk size is fixed by requiring that the disk mass fraction (4.6\% of the total mass) is equal to the disk angular momentum.  This results in a radial scale length for the exponential disk of $4.1~\rm kpc$.  In addition, a specified fraction $f_{\rm gas}$ of the mass of the disk is in a gaseous component, where $f_{\rm gas}=0.2$ for the fiducial model.  The mass fraction of the extended HI disk relative to the total gas mass is equal to 0.3, and its scale length is three times that of the exponential disk of gas and stars.  In dimensional terms, this model at onset has stellar, gas, and halo masses of $3.5 \times 10^{10} M_{\odot}, 1.25 \times 10^{10} M_{\odot}$ and $1.44 \times 10^{12} M_{\odot}$ respectively.

The dark sub-halo in the fiducial model that tidally interacts with the primary galaxy is also initialized with a Hernquist profile.
The concentration of the dark sub-halo is scaled by the mass of the sub-halo using equation 9 of Maccio et al. (2008).  Cosmological dark-matter only simulations find little variation of the spin of halos with mass (Bullock et al. 2001; Maccio et al. 2008), so we leave this parameter fixed.  The circular velocity of the sub-halos is scaled as $M^{1/3}$ as usual.

Cosmological dark matter only simulations indicate the prevalence of parabolic orbits (Khochfar \& Burkert 2006) for CDM sub-structures.  Hence, we are motivated to set the dark sub-halos on a parabolic orbit around the primary galaxy.  We take the starting separation to be $100~\rm kpc$, which is of order the virial radius of the Milky Way, and vary the mass, pericentric distance, and inclination of the sub-halos.  For a given configuration of the primary galaxy, these three parameters (mass, pericentric distance, and orbital inclination) are the only parameters that can be varied.  Table 1 lists the parameters of a representative sub-set of the simulations we have performed.   We adopt the nomenclature from the merger simulation survey paper by Cox et al. (2006), where the angles ($\theta,\phi$) of the angular momentum vectors of the primary (1) and secondary (2) galaxy are defined relative to the orbital plane.  As in Cox et al. (2006), the ``h'' inclination corresponds to the angular momentum vectors being co-planar, i.e., a prograde-prograde interaction, with $\theta_{1}=0=\phi_{1}=\theta_{2}=\phi_{2}=0$, where 1 and 2 refer to the primary galaxy and the sub-halo respectively.  The ``md'' inclination corresponds to $\theta_{1}=\phi_{1}=0,\theta_{2}=90,\phi_{2}=0$.  The simulations are labeled in Table 1 by the mass of the satellite and the pericentric approach distance, for instance the 100R5 simulation refers to a 1:100 satellite interacting with the primary galaxy, having a pericentric approach distance of 5 kpc; if the mergers are not co-planar their label (and the inclination listed in Table 1) indicates that.  Table 1 also includes one case with a gaseous component in the perturber (with $f_{\rm gas}=0.2$ for the primary galaxy and the perturber), the 100R15bary case.

GADGET-2 uses a bulk artificial viscosity which is the shear-reduced version (Balsara 1995; Steinmetz 1996) of the standard Monaghan and Gingold (1983) artificial viscosity.  Here, we adopt a fiducial value of 0.75 for the bulk artificial viscosity parameter.  The version of GADGET-2 that we employ uses a sub-resolution model for energy injection from supernovae that is proportional to the star
formation rate (Springel \& Hernquist 2003).  The effective equation of state (EQS) for this two-component gas (Eq. 24 in Springel et al. 2001) is varied from 0-1.

Star formation is modeled using a density dependent Kennicutt-Schmidt algorithm.  The star formation rate is given by:

\begin{equation}
\frac{d\rho_{\star}}{dt}=(1-\beta)\frac{\rho}{t_{\star}}  \;,
\end{equation}
where $\rho$ is the total gas density, $\beta=0.106$ for a Salpeter (1955) IMF with slope -1.35 and upper and lower limits of 40 and 0.1 $M_{\odot}$ respectively, and the quantity $t_{\star}$ is given by:

\begin{equation}
t_{\star}(\rho)=t_{0}^{\star}\left(\frac{\rho}{\rho_{\rm th}}\right)^{-1/2} \;,
\end{equation}
where $\rho_{\rm th}$ is fixed by the multiphase model and is the
threshold density above which the gas is assumed to be thermally
unstable to the onset of a two-phase medium composed of cold star
forming clouds embedded in a hot, pressure-confining phase (SH03).  We
adopt $\beta=0.1$ in this paper; SH03 previously found that the star
formation is not sensitively dependent on this choice.  We take $t_{0}^{\star}$, the gas consumption timescale, to be 4.5 Gyr which is an intermediate value between that adopted by SH03
and Springel et al. (2005) and agrees reasonably well (to within a
factor of 2-4) with gas consumption time scales cited by Leroy et al.
(2009).  

We have performed extensive convergence studies to verify that our results are converged.  We note in summary that varying the particle number by a factor of 3 on either side of our fiducial model values and softening length $\epsilon$ (where $\epsilon$ is varied as $N^{1/3}$ as usual), results in a difference in radially varying Fourier amplitudes less than a factor of 1.3, while integrated quantities such as the mass and angular momentum are converged to better than a few percent.

\begin{table}
\begin{minipage}{5mm}
{\small
  \begin{tabular}{@{}lcccccc}
  \hline

Sim & $f_{\rm gas}$  &  $V_{200}$ & EQS & $R_{\rm peri}$ & incl. & Fig.2c \\

\hline

100E0R5 &  0.2   & 160  & 0     & 5    & h  & green, circle \\
100R5  &  0.2   & 160  & 0.25  & 5    & h  & green, cross \\
100v4R5  & 0.2  & 165  & 0.25  & 5    & h  & green, asterisk \\
100v40.1R5 & 0.1  & 165 & 0.25  & 5    & h  & green, diamond \\
100E0R5md    &  0.2   & 160  & 0     & 5    & md  & green, triangle \\
100R15 & 0.2  & 160  & 0.25  & 15   & h  & black, circle \\
100R15d & 0.2  & 160  & 0.25  & 15    & d  & black, asterisk  \\
100R15bary & 0.2  & 160  & 0.25  & 15 & h  & black, diamond \\
300R15   & 0.2  & 160  & 0.25  & 15  & h & blue, circle\\
50R15 & 0.2  & 160  & 0.25  & 15     & h  & orange, circle\\
10R15 & 0.2  & 160  & 0.25  & 15     & h   & red, circle \\
10R32 & 0.2  & 160  & 0.25  & 32     & h   & red, asterisk \\

\hline
\end{tabular}}
\end{minipage}
\end{table}

\section{Results}

We may expect that passing dark sub-halos will gravitationally sculpt gas disks on the outskirts of galaxies, leaving tidal imprints of their passage.  We now describe the analysis of our set of simulations ($\sim 50$) of which we have selected 12 in Table 1 to illustrate the parameters varied.  For all simulations and all simulation snapshots, we calculate the Fourier modes of the gas surface density as a function of time:
\begin{equation}
a_{m}(r,t)=\frac{1}{2\pi}\int_{0}^{2\pi} \Sigma(r,\phi,t)e^{-im\phi}d\phi  \;.
\end{equation}   
We compare this azimuthally integrated quantity to the Fourier modes of the raw HI data.  Due to the velocity smearing in the HI data that becomes more problematic at large radii ($r \ga 30~\rm kpc$), and confusion and distance determination problems interior to the solar circle, we restrict our analysis to radii greater than 10 kpc and less than 25 kpc.  Our determination of the best-fit simulation snapshot derives from the calculation of the residuals of the $m=0-4$ modes of the data and the simulations.  We restrict our analysis to these modes, as most of the power is in these modes.  Specifically, most of the power is in the $m=1$ mode, and we analyze those modes which have median values of $\left[a_{m}'(r)/a_{1}(r)\right]\ga 0.4$ in the range $10 \la r \la 25~\rm kpc$, where $a_{m}'(r)=a_{m}(r)/a_{0}(r)$.  This corresponds to the first four Fourier modes.  The residuals for a given simulation are calculated as follows:  $S=\sum_{r} \left[a_{1,D}'(r)-a_{1,S}'(r,t)\right]^{2}\times\left(a_{1,D}'(r)\right)^{2}+\left[a_{2,D}'(r)-a_{2,S}'(r,t)\right]^{2}\times\left(a_{2,D}'(r)\right)^{2}+\left[a_{3,D}'(r)-a_{3,S}'(r,t)\right]^{2}\times\left(a_{3,D}'(r)\right)^{2}+\left[a_{4,D}'(r)-a_{4,S}'(r,t)\right]^{2}\times\left(a_{4,D}'(r)\right)^{2}$.  Here, $a_{m,D}'$ and $a_{m,S}'$ denote the modes for the data (D) and the simulation (S) respectively.  The best-fit time snapshot is that which minimizes $S$ for a given simulation.  The entire simulation set is searched accordingly.  The quantities $S_{1}$ and $S_{1-4}$ will refer to the variances of the $m=1$ and $m=1-4$ modes respectively.

\begin{figure}
\begin{center}
\includegraphics[scale=0.25]{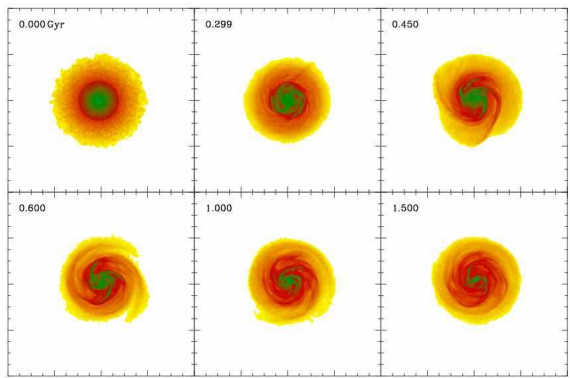}          
\caption{Projected gas density images for the 100E0R5 simulation out to 1.5 Gyr.  The length of the box is 40 kpc.  The t=0.29 Gyr time-snap corresponds to pericentric distance $t(R_{\rm peri})$, and the t=0.6 Gyr time-snap corresponds to the best-fit to the Fourier modes of the HI data.}
\end{center}
\end{figure}

Figure 1 shows images of the gas surface density of the best-fit
simulation, starting from the initial conditions out to 1.5 Gyr.
Pericentric approach occurs at $t=0.29~\rm Gyr=t(R_{\rm peri})$, and
the maximal visual response occurs about one dynamical time after pericentric approach, i.e., ($t(R_{\rm peri})+\sim 250~\rm Myr \sim 0.6~\rm Gyr$).  These tidal disturbances in the gas disk are of a very different character from spiral density waves.  The disturbances have a wavelength that is increasing outwards, while the wavelength of spiral density waves decreases outwards (Shu 1983).  This is therefore a very different response than that of the gas to the stellar spiral arms, as has been studied in previous works, either through a self-consistent calculation of the stars and gas (Chakrabarti 2009), or through an assumed potential for the stellar spiral potential forcing the gas (Chakrabarti, Laughlin \& Shu 2003; Shetty \& Ostriker 2006).  The tidal response is in principle akin to ``flapping a bedsheet'', where the wavelength of the disturbances increases outwards.    

\begin{figure}
\begin{center}
\includegraphics[scale=0.4]{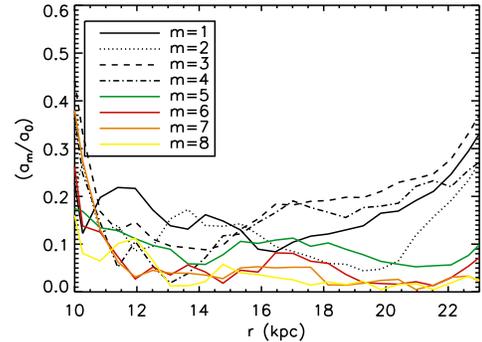}
\includegraphics[scale=0.4]{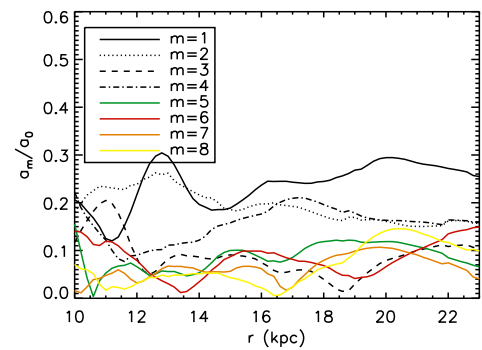}          
\includegraphics[scale=0.4]{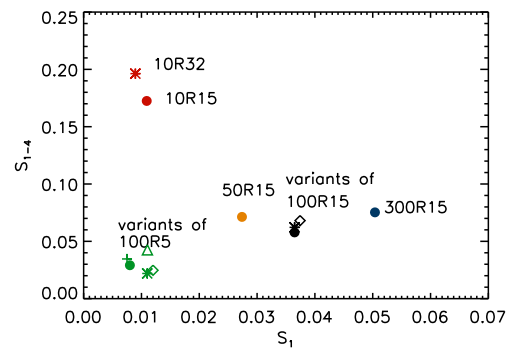}
\caption{(a) Fourier modes of the gas surface density, normalized to $m=0$, at $t=0.6~\rm Gyr$ of 100E0R5, which is the snapshot that best fits the planar disturbances, (b) Fourier modes of the raw HI data, and (c) $S_{1}$ vs $S_{1-4}$ for the simulations in the table; best-fit case (the one that minimizes the distance to the origin) is shown in green filled circle; see table for labels}
\end{center}
\end{figure}

Figure 2(a) depicts the $m=1-8$ Fourier modes of the gas surface density at the best-fit snapshot, which is $t=0.6~\rm Gyr$ for the 100E0R5 simulation.  Of note is the rise in the $m=1$ power at $r\sim 13~\rm kpc$ in the Fourier mode of the raw HI data as shown in Figure 2b; searching for the minimum of the residuals also selects a simulation snapshot ($t=0.6~\rm Gyr$) that evinces a similar azimuthal asymmetry.  Figure 1 of Levine, Blitz \& Heiles (2006) displays a similar azimuthal asymmetry in the image itself.  There is a considerable amount of power even out to high m-modes, a feature that is seen across all simulations of minor mergers or tidal interactions.  We also note that other properties of the simulated disk at the best-fit snapshot, such as the star formation rate, which is $\sim 4~M_{\odot}/\rm yr$, and stellar mass also match estimates for the Milky Way reasonably (McKee \& Williams 1997).

We have examined the sensitivity of our results to the initial conditions.  While the detailed shape of the planar disturbances varies between simulations like 100R15 and 100R15d (with the ``h'' co-planar inclination having the largest planar disturbances) cases, or between 100R5 as performed with EQS=0 (isothermal) or with EQS=0.25 (with the isothermal equation of state producing somewhat larger disturbances), the inference of a 1:100 mass ratio perturber is robust to these variances.  In other words, varying the equation of state, gas fraction (to within the range of typical Milky Way values), or orbital inclination will not favor a significantly larger or smaller perturber.  Figure 2(c) demonstrates this point by plotting $S_{1-4}$ vs $S_{1}$.  While changes in initial conditions will lead to changes in the Fourier modes, the primary determinants of the Fourier modes are the mass of the perturber and the pericentric distance.  The best-fit simulation (100E0R5) is that which minimizes the distance to the origin, i.e., the one with the lowest $S_{1}$ and $S_{1-4}$.  The errors in the data are less than 20 \%; statistical errors are negligible, systematic uncertainties are thought to be less than 20 \% in the data set (Brandt \& Blitz 1993; Levine et al. 2006).  As such, differences in variances greater than a factor of 2 should be statistically significant.  In a future work, we perform a more generic investigation of distortions in the HI disks of bright spiral galaxies, including the effects of cold flows (Keres \& Hernquist 2009; Brooks et al. 2009).  For perturber gas fractions $\sim 0.2$ (as in the 100R15bary case), the differences between dark sub-halos and inflowing gaseous clumps may be somewhat discernible, but it may not be possible to clearly discriminate due to differences in initial conditions which does lead to some spread in the Fourier modes (as shown by the green points).

We note briefly that the mass-weighted half thickness averaged over azimuth
of the best-fit simulation matches the data approximately, within a factor of $\sim 2$.  The thickness in the $0 < \phi < 180$ regime is higher than in $180 < \phi < 360$, as is the case in the data (Levine et al. 2006a).  The distribution of thickness values in these two azimuthal regimes is not as skewed as it is in the data.  We speculate that the warp of our galaxy may also have a contribution from a non-spherical potential for the dark matter halo.  Such asymmetries (i.e., a $\theta$ component of the halo of our galaxy) will primarily affect the development of warps or vertical disturbances (as the torque due to such a $\theta$ component in the potential primarily affects the angular momentum in the $z$ direction), but not the planar disturbances.  Such an analysis is common in planetary disks (Chiang et al. 2009), where the vertical disturbances are largely decoupled from the planar disturbances.  Weinberg \& Blitz (2006) examined the warp of the galaxy via analysis of the Fourier modes of the vertical height distribution, and found agreement with the data at a factor of $\sim 2$ level.  It appears likely that both perturbers and a non-spherical halo potential may play a role in determining the vertical disturbances of the gas disk.  

\begin{figure}
\begin{center}
\includegraphics[scale=0.25]{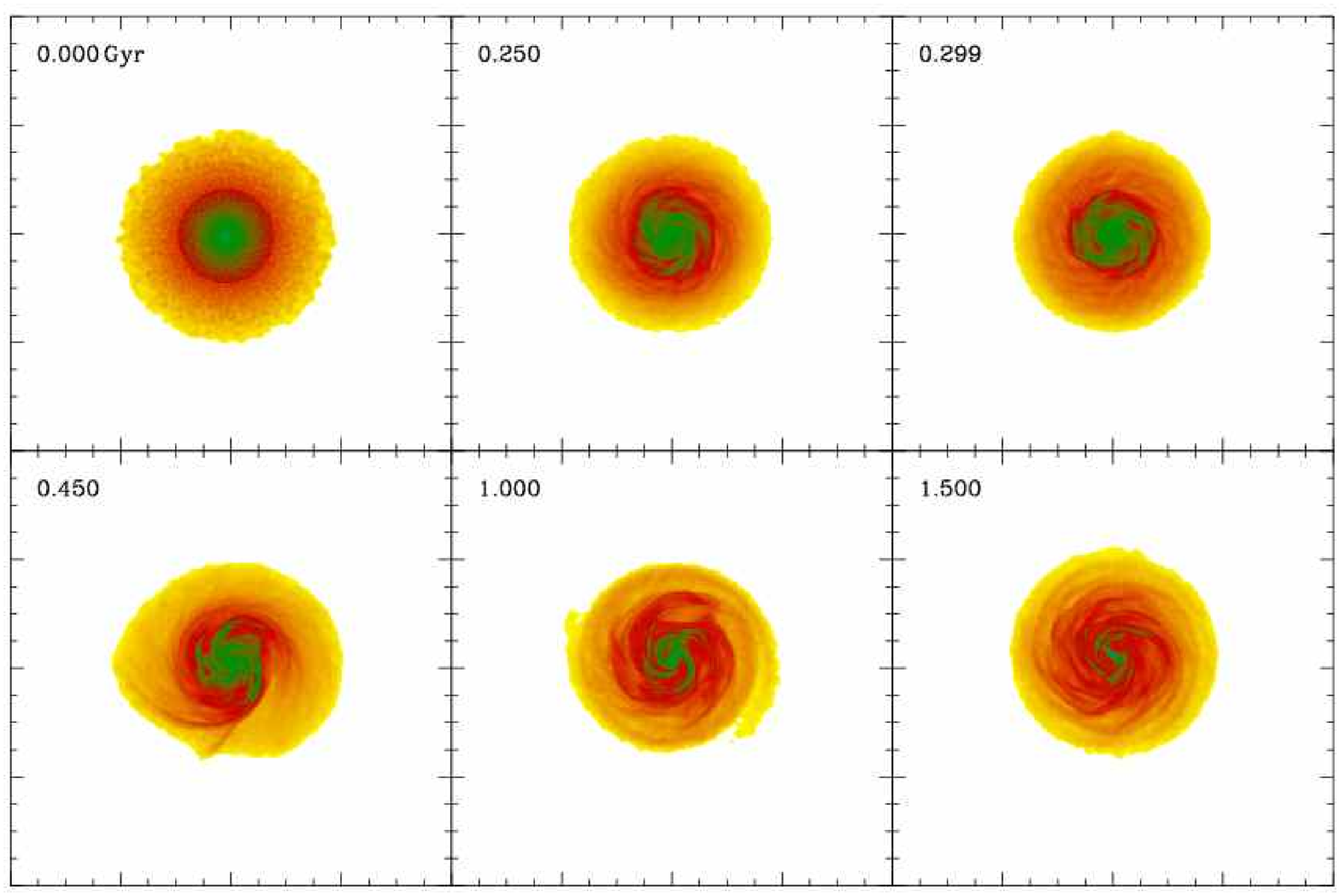}
\includegraphics[scale=0.25]{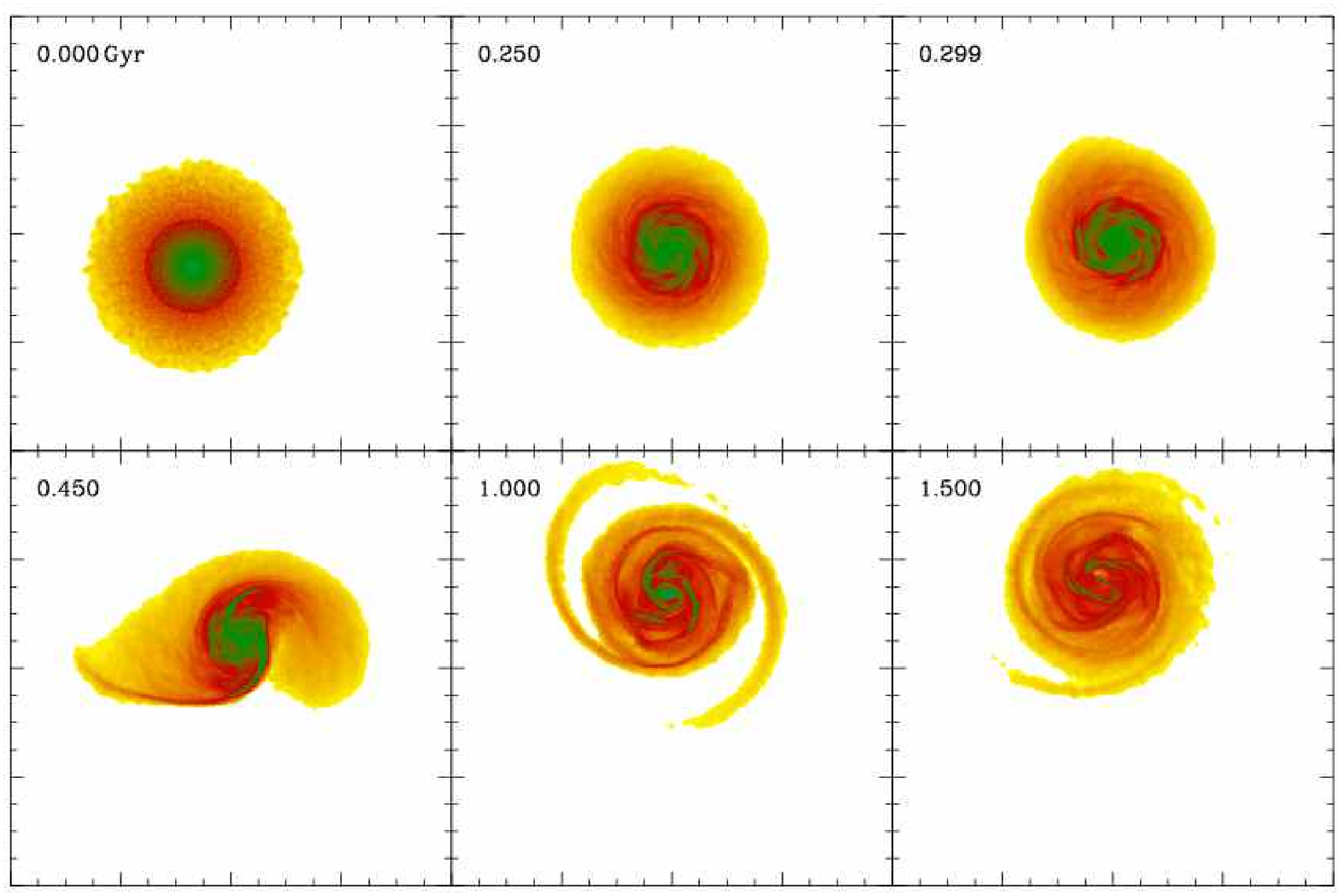}
\caption{Projected gas density images for the 100R15 and 10R32 simulations out to 1.5 Gyr.  These satellites by construction exert the same tidal force at $R_{\rm peri}$.  The length of the box is 40 kpc. (a) Interaction of a Milky Way like disk with 1:100 satellite having $R_{\rm peri}=15 ~\rm kpc$; (b) This is a 1:10 merger at the equivalent tidal distance as the 100R15 simulation, i.e., at $R_{\rm peri}$ both of these satellites exert the same tidal force by construction, but elicit a different response in the primary galaxy over time}
\end{center}
\end{figure}

\subsection{Tidal Approach of Mass Determination:  Breaking Degeneracy Between M \& R}

To infer the mass of perturbing sub-halos (rather than the quantity
$M_{\rm sat}/R_{\rm peri}^{3}$ to which the tidal force is
proportional), it is necessary to determine the conditions under which
one can discriminate between varying mass ratio mergers at equivalent
tidal distances.  We present the essential point and demonstration of
the idea in this paper, and defer the details to a longer paper.  In
Figure 3 (a-b), we show the gas density images as a function of time
for two simulations where the satellites exert the same tidal force
at $R_{\rm peri}$ by construction.  Figure 3(a) depicts the gas
density response of a simulation of the Milky Way undergoing a tidal
interaction with a sub-halo with a mass that is one-hundredth that of
the primary galaxy, and a pericentric approach distance of 15 kpc.  Figure 3(b) depicts the response of the galaxy when it undergoes a tidal interaction with a satellite of mass one-tenth that of the primary galaxy with pericentric approach distance of $R_{2}=(M_{2}/M_{1})^{1/3}R_{1}$.  In other words, we have placed a satellite $M_{2}$ (a 1:10 mass ratio satellite) at the equivalent tidal distance as $M_{1}$ (a 1:100 mass ratio satellite).  It is immediately clear from comparison of Figures 3 (a) and (b) that the response of a gas disk that experiences a tidal interaction with a satellite $M_{2}$ placed at the equivalent tidal distance as $M_{1}$ is significantly different (even though the tidal force at $t(R_{\rm peri})$ is exactly the same by construction).  

The basic reason for this is that the satellite begins to exert an
influence on the gas disk at a location $R_{0}(M_{s})$ (where $M_{s}$
is the mass of the satellite).  On a parabolic orbit, the influence of
a satellite is not limited to occuring at the distance of closest
approach (although the tidal torque it exerts is largest at that
point).   We denote $R_{0}(M_{s})$ as the location at which a
satellite (of a given mass $M_{s}$) first begins to influence the gas
disk of a galaxy in so far as inducing tidal torques that produce
velocities in excess of the sound speed.  We find that we can break
the degeneracy between the mass of the satellite and the cube of the
distance if the time it takes for the satellite to move from $R_{0}$
to $R_{\rm peri}$ is larger than the time it takes the gas to shock.
That is, $t(R_{0})-t(R_{\rm peri}) \ga t_{\rm shock}$, where $t_{\rm shock}$ is of order the time it takes for the gas to cross spiral arms.  It becomes increasingly harder to determine the mass (rather than the quantity $M_{s}/R_{\rm peri}^{3}$) for interactions with satellites of very small mass, as their $R_{0}$ will approach $R_{\rm peri}$, resulting in a (nearly) impulsive tidal effect.  This effect (i.e., the buildup of a non-linear density response in the gas disk) occurs for satellites that are sufficiently massive such that $R_{0}(M_{s}) \ga R_{\rm peri}$.  We also find that satellites of mass $M_{2}$ and pericentric approach distance of $R_{2}$ (as defined above), but on a polar or tilted orbits (as opposed to co-planar as in Figure 3 (a) and (b)) do not show a markedly different response.  

We also note that the analysis of the proper motions of the LMC (Kallivayalil et al. 2006) indicate that the LMC is just approaching 50 kpc; this result holds even when the depth of the potential well is varied (Shattow \& Loeb 2009).  A 1:100 perturber (such as the LMC) can produce the level of power in the Fourier modes as are seen in the data if $R_{\rm peri} \sim 5-10~\rm kpc$.   However, this response occurs a dynamical time after $t(R_{\rm peri})$.  The proper motions indicate that the LMC is just approaching 50 kpc, not receding after first passage.  Therefore, the timing of the orbit that follows from the proper motion analysis precludes the LMC as the culprit for producing these disturbances, as a 1:100 perturber just approaching 50 kpc will not raise appreciable tides on the Milky Way disk.  The Sagittarius dwarf is nearby (16 kpc from the galactic center), and modeling of its tidal debris suggests that it may have initially been more massive (Johnston et al. 1999), with a first impact at $\sim 40~\rm kpc$.  Proper motions of Sgr (Ibata et al. 1997) indicate that it is currently moving towards the plane.  If the impact parameter of the earlier passage has been overestimated (i.e., if it is $\sim 5~\rm kpc$ as opposed to $\sim 40~\rm kpc$), then it may be responsible for the observed perturbations in the HI disk, but this would require significant revision of the orbit calculations for Sgr.

Our analysis indicates that the perturber responsible for the tidal imprints on the HI disk is moving away from the galaxy, at a current distance of 90 kpc from the galactic center, and is massive, with a mass that is one-hundredth of the Milky Way.  Perturbers with mass ratios less than 1:300 or greater than 1:50 cannot produce the observed perturbations in the HI disk; the favored pericentric approach distance is of order the scale length of the Milky Way disk.

\section{Conclusion}

$\bullet$  By analysis of the observed Fourier modes of the HI disk of
our galaxy and our set of simulations, we infer that the perturbations
in the gas disk are due to a dark sub-halo with a mass one-hundredth of the Milky Way, with a pericentric distance of 5 kpc.  The best-fit time corresponds to about a dynamical time after
closest approach, when the perturber is 90 kpc from the center of the galaxy.  We have also shown that variations in initial conditions would not significantly alter our inference of the perturber mass or pericentric distance.

$\bullet$ We have illustrated a fundamental property of parabolic orbits that allows us to determine the mass of the perturber directly (rather than the tidal force).  We show that if the time for the perturber to move from $R_{0}$ (the closest distance at which the perturber starts to signficantly affect the galaxy) to $R_{\rm peri}$ is of order the time for the gas to shock, then one can break the degeneracy between the mass of the perturber and the pericentric distance in the tidal force.  

$\bullet$  We have presented a new method, which we term the $\it{tidal~approach~for~mass~determination}$ that can be employed to determine the masses of dark sub-halos from the tidal gravitational imprints they leave on galactic gas disks.  A corollary of this method is that observations of quiescent galaxies which do not evince perturbations in the outer disk imply a lower-bound to the perturber mass for a given surface density of the galactic disk.

\section*{Acknowledgements}

We thank Frank Shu, Carl Heiles, T.J. Cox, Chris McKee, Marc Davis, Avi Loeb, Eugene Chiang, Eliot Quataert and especially Phil Chang for helpful discussions, and Evan Levine for providing the observational data.  We also thank the referee, Fabio Governato, for his helpful comments and feedback.  SC is supported by a UC Presidential Fellowship, and LB is partially supported by funding from NSF grant AST-0838258, and the Paul G Allen Family Foundation.  


\end{document}